\documentclass[aps,twocolumn,prc,superscriptaddress,showpacs,nofootinbib,floatfix,amssymb,amsfonts,amsmath]{revtex4-1}

\usepackage{graphicx}
\usepackage{dcolumn}
\usepackage{bm}

\usepackage{amsmath}    
\usepackage{amsfonts}   
\usepackage{amssymb}
\usepackage{graphicx}   
\begin{document}
\title{Superheavy nuclei in microscopic collective Hamiltonian 
approach: the impact of beyond mean field correlations on the 
ground state and fission properties. }

\author{Z.\ Shi}
\affiliation{School of Physics and Nuclear Energy Engineering,
             Beihang University, Beijing 100191, China}

\author{A.\ V.\ Afanasjev}
\affiliation{Department of Physics and Astronomy, Mississippi
State University, MS 39762, USA}
\affiliation{Yukawa Institute for Theoretical Physics, Kyoto University,
             Kyoto 606-8502, Japan}

\author{Z.\ P.\ Li}
\affiliation{School of Physical Science and Technology, Southwest
University, Chongqing 400715, China}

\author{J.\ Meng}
\affiliation{State Key Laboratory of Nuclear Physics and Technology,
             School of Physics, Peking University, Beijing 100871,
             China}
\affiliation{Yukawa Institute for Theoretical Physics, Kyoto University,
             Kyoto 606-8502, Japan}

\date{\today}

\begin{abstract}
The impact of beyond mean field effects on the ground state and fission
properties of superheavy nuclei has been investigated in a five-dimensional
collective Hamiltonian based on covariant density functional theory.
The inclusion of dynamical correlations reduces
the impact of the $Z=120$ shell closure and induces substantial collectivity
for the majority of the $Z=120$ nuclei which otherwise are spherical at the
mean field level (as seen in the calculations with the PC-PK1 functional).
Thus, they lead to a substantial convergence of the predictions of the
functionals DD-PC1 and PC-PK1 which are different at the mean field level.
On the
contrary, the predictions of these two functionals remain distinctly
different for the $N=184$ nuclei even when dynamical correlations
are included. These nuclei are mostly spherical (oblate) in the calculations
with PC-PK1 (DD-PC1). 
 Our calculations for the first time reveal significant impact of 
dynamical correlations on the heights of inner fission barriers of 
superheavy nuclei with soft potential energy surfaces, the 
minimum of which at the mean field level is located at spherical
shape. 
These correlations affect the fission barriers of the nuclei, which
are deformed in the ground state at the mean field level, to a lesser degree.
\end{abstract}


\maketitle

\section{Introduction}

  One of the most active sub-fields of low-energy nuclear physics
is the investigation of superheavy elements (SHE) \cite{OU.15}. At
present, the nuclear chart extends up to the element Og with proton
number $Z=118$ \cite{Z=117-118-year2012}. However, the experimental
difficulties in the studies of SHEs at this extreme of the proton
number are enormous: the experiments lasting several months typically
provide only few events \cite{OU.15}. New facilities such as
Superheavy Element Factory in Dubna, Russia \cite{OD.16} will allow
to observe substantially more events at presently available $Z$
values and hopefully to extend the nuclear chart to higher $Z$
values.

  In addition to experimental challenges, there are substantial
theoretical uncertainties related to the predictions of the position
of the center of the island of stability of superheavy elements
\cite{MN.94,BRRMG.99,BNR.01,AANR.15} and their fission properties
\cite{BKRRSW.15,AARR.17}. Different models locate this center at
different particle numbers. For example, the microscopic+macroscopic
(MM) models put it at $Z=114, N=184$ \cite{MN.94,NNSSWGM.68,PS.91}.
Most of the Skyrme energy density functionals (SEDF) place it at
$Z=126, N=184$ \cite{BRRMG.99,BNR.01}. However, there are also some
SEDFs which predict large shell gap at $Z=120$ \cite{BRRMG.99}.

   Note that the number of these predictions was obtained in the
calculations restricted to spherical shape. The danger of this
restriction has recently been illustrated in the covariant density
functional theoretical (CDFT \cite{VALR.05,MTZZLG.06,RDFNS.16}) 
study of Ref.\ \cite{AANR.15} based on axial relativistic 
Hartree-Bogoliubov (RHB) calculations.
Earlier CDFT studies 
\cite{BRRMG.99,BNR.01,Rutz_PhysRevC.56.238.1997,A250,ZMZGT.05}
restricted to spherical shape almost always indicated $Z=120, N=172$ as
the center of the island of stability of SHEs. However, the inclusion
of deformation has drastically changed this situation: it was found that
the impact of the $N = 172$ spherical shell gap on the structure of SHE is very
limited. Similar to non-relativistic functionals, some covariant functionals
predict the important role played by the spherical $N = 184$ gap. For these
functionals (NL3* \cite{NL3*}, DD-ME2 \cite{DD-ME2}, and PC-PK1 \cite{PC-PK1})
there is a band of spherical nuclei along and near the $Z = 120$ and $N = 184$
lines. However, for other functionals (DD-PC1 \cite{DD-PC1} and DD-ME$\delta$
\cite{DD-MEdelta}) oblate shapes dominate at and in the vicinity of
these lines. Available experimental data (which do not extend up to the
$Z=120$ and $N=184$ lines) are described with comparable accuracy in the
calculations with these functionals which does not allow to discriminate
between these predictions. Note that all these functionals are globally
tested \cite{AANR.15,AARR.17,AARR.14} and only DD-ME$\delta$ is not
recommended for the nuclei beyond lead region based on the studies of
inner fission barriers \cite{AARR.17} and octupole deformed nuclei
\cite{AAR.16}.

\begin{figure*}[htb]
\centering
\includegraphics[scale=0.46,angle=0]{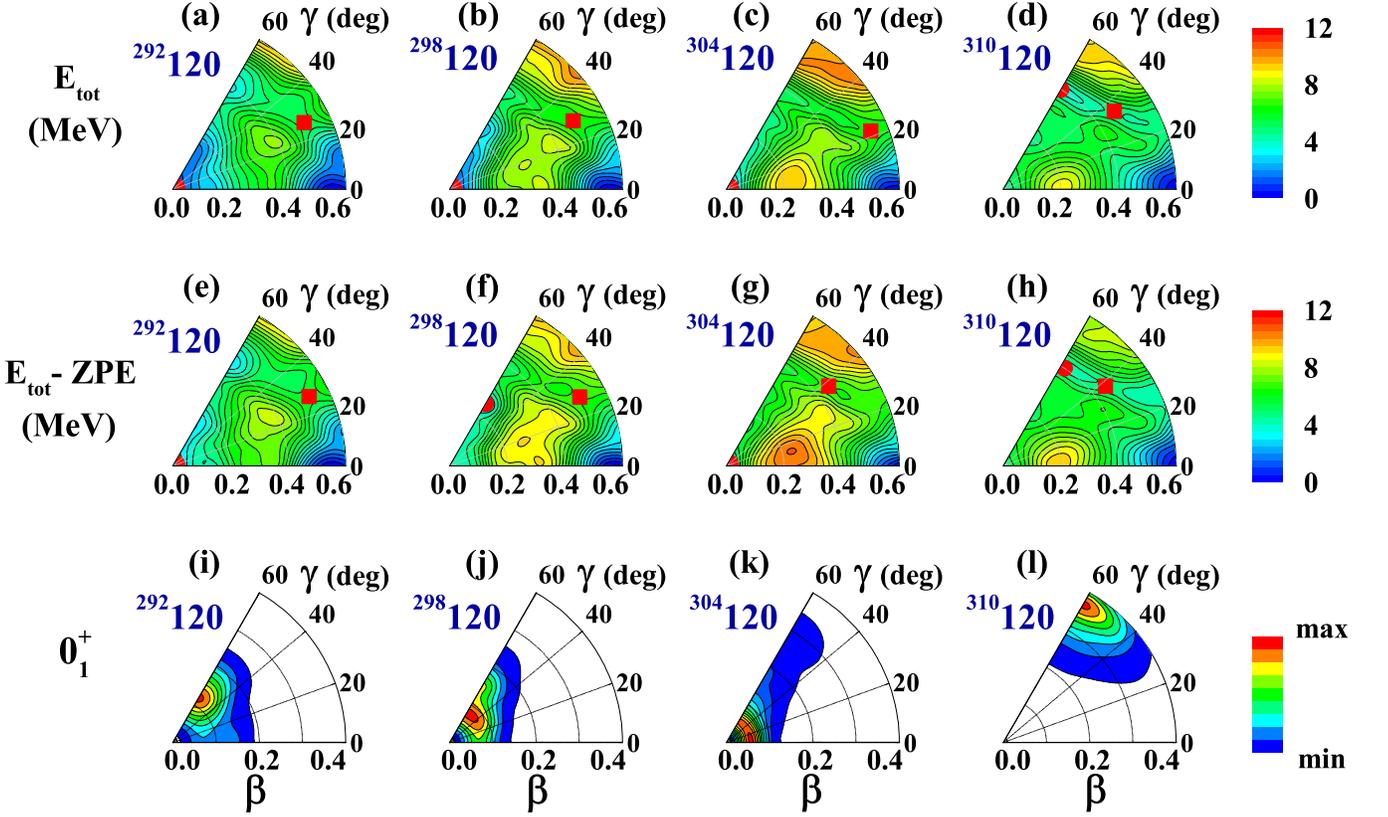}
\caption{(Color online) Potential energy surfaces (top panels),
collective energy surfaces (CES) with zero-point-energy (ZPE) taken into
account (middle panels) and probability density distributions (in
arbitrary units) in the $\beta$-$\gamma$ plane for the $0_{1}^+$
states (bottom panels) of selected nuclei in  $Z=120$ isotopic chain.
The results are obtained with PC-PK1 CEDF. The energy difference
between two neighboring equipotential lines is equal to 0.5 MeV.
The minima and saddles in top and middle panels are shown by
circles and squares, respectively.
}
\label{pes-z=120-pcpk1}
\end{figure*}

  The results obtained in Ref.\ \cite{AANR.15} on the structure
of the ground states of superheavy nuclei could be further modified.
This is because the potential energy surfaces of many nuclei along
the $Z=120$ and $N=184$ lines are soft in quadrupole deformation
(see Figs. 3 and 4 of Ref.\ \cite{AANR.15} and Fig.\ \ref{pes-z=120-pcpk1}
in the present manuscript). For such transitional nuclei the correlations
beyond mean field may substantially modify the physical situation, for
example, by creating deformed ground state instead of spherical one
at the mean field level. However, this issue has not been investigated
before since the studies of SHEs are almost always done on the mean field
level. Only in Ref.\ \cite{PNLV.12} the beyond mean field effects have
been taken into account for the ground states of several SHE located in
the $\alpha$-decay chains of the $^{298,300}$120 nuclei in the relativistic
calculations based on the DD-PC1 functional.

  Another question of interest is the impact of dynamical
correlations on the fission barrier heights. So far the majority
of the fission barrier calculations have been performed at the
mean field level (see Refs.\ \cite{BKRRSW.15,AARR.17,MSI.09,AAR.10,AAR.12,SR.16}
and references therein). 
  Substantial differences in the predictions of inner fission barrier heights 
for SHE  existing between different non-relativistic  and relativistic models and 
between different covariant energy density functionals in the CDFT calculations 
are summarized  in Figs. 12 and 10 of Ref.\ \cite{AARR.17}, respectively.  The CDFT 
predictions for the inner fission barrier heights of SHE are located at the lower end 
of the range of predictions of all considered models/functionals in these figures.

The importance of dynamical correlations in triaxial calculations has been studied only 
for few actinide \cite{DGGL.06,SDNSB.14,ZLNVZ.16,BMIQ.17} and light superheavy
\cite{GSPS.99} nuclei. However, the impact of dynamical correlations on fission 
barriers of SHE has not been studied in a relativistic framework.
Contrary to the actinides in which the ground state is prolate
deformed, the situation for superheavy nuclei in the vicinity of the
$Z=120$ and $N=184$ lines is different since such nuclei have either
spherical or oblate deformation in the ground state and are transitional
in nature \cite{AANR.15}. Considering significant impact of the fission
barriers on the stability of superheavy nuclei, it is necessary to evaluate
the impact of dynamical correlations on their heights.

  The present manuscript aims  at the investigation of the impact of
dynamical correlations on the ground state and fission properties of
superheavy nuclei along the $Z=120$ isotopic chain (with $N=172-190$), 
$N=174$ (with $Z=108-124$) and $N=184$ (with $Z=112-122$) isotonic 
chains.  The 
calculations are performed within five-dimensional collective Hamiltonian 
(5DCH) approach  \cite{LNVMLR.09,NLVPMR.09,NVR.11} based on CDFT which 
has been extremely successful in the description of many physical phenomena
\cite{LNVM.09,LNVRM.10,LYVNCM.11,LLXYM.12,WSCZS.16,QCLNV.17,SL.18,SCZ.18}.

  They are carried out with two covariant energy density functionals 
(CEDFs), namely, PC-PK1 \cite{PC-PK1} and DD-PC1 \cite{DD-PC1},
representing two extremes of the predictions for superheavy nuclei in the
CDFT. PC-PK1 predicts the bands of spherical nuclei along $Z=120$ and $N=184$
\cite{AANR.15} suggesting that the $^{304}$120 nucleus may be considered as
doubly magic. On the contrary, the nuclei along these lines and beyond are
oblate in the calculations with DD-PC1 \cite{AANR.15}. Note that these two
functionals provide the best description of experimental data in actinides 
and superheavy nuclei among 5 employed in Ref.\ \cite{AANR.15} state-of-the-art
CEDFs.

In Sec.~\ref{sec1} we present a short outline of  theoretical framework for the 5DCH 
approach based on CDFT.  The systematics of collective potential energy surfaces, 
deformations, low-energy spectra, and fission barriers are discussed in Sec.~\ref{sec2}. 
Section \ref{sec3} summarizes the principal results.

\section{Theoretical Framework}\label{sec1}

The 5DCH that describes the nuclear excitations of quadrupole vibration and rotation is 
expressed in terms of two deformation parameters $\beta$ and $\gamma$ and three Euler 
angles $(\phi,\theta,\psi)\equiv \Omega$ \cite{LNVMLR.09,NLVPMR.09,NVR.11},
%
\begin{equation}
    \hat H_{\rm coll}(\beta,\gamma,\Omega)=\hat T_{\rm vib}(\beta,\gamma)
    +\hat T_{\rm rot}(\beta,\gamma,\Omega)+V_{\rm coll}(\beta,\gamma). \\
    \label{eq1}
\end{equation}
%
The three terms in $\hat H_{\rm coll}(\beta,\gamma,\Omega)$ are the vibrational kinetic energy
%
\begin{eqnarray}
\hat{T}_{\rm vib}=&-\frac{\hbar^2}{2\sqrt{wr}}\left\{\frac{1}{\beta^4}\left[\frac{\partial}{\partial \beta}\sqrt{\frac{r}{w}}\beta^4B_{\gamma\gamma}\frac{\partial}{\partial \beta}
-\frac{\partial}{\partial\beta}\sqrt{\frac{r}{w}}\beta^3 B_{\beta\gamma}\frac{\partial}{\partial\gamma}\right] \right.
\nonumber \\ 
&+\frac{1}{\beta \sin 3\gamma}\left[-\frac{\partial}{\partial\gamma}
\sqrt{\frac{r}{w}}\sin3\gamma B_{\beta\gamma}\frac{\partial}{\partial\beta} \right. 
\nonumber \\ 
&\left.\left.+\frac{1}{\beta}\frac{\partial}{\partial\gamma}\sqrt{\frac{r}{w}}
\sin 3\gamma B_{\beta\beta}\frac{\partial}{\partial\gamma}\right]\right\},
\label{eqnarray}
\end{eqnarray}
%
the rotational kinetic energy
%
\begin{equation}
   \hat T_{\rm rot}=\frac{1}{2}\sum^3_{k=1}\frac{\hat J_k^2}{\mathcal I_k},
   \label{eq3}
\end{equation}
%
and the collective potential $V_{\rm coll}$, respectively. Here, $\hat J_k$ denote the components of the total angular momentum in the body-fixed frame, and both the mass parameters $B_{\beta\beta}$, $B_{\beta\gamma}$, $B_{\gamma\gamma}$ and the moments of inertia $\mathcal I_k$ depend on the quadrupole deformation variables $\beta$ and $\gamma$. Two additional quantities that appear in the $\hat{T}_{\rm vib}$ term, $r = B_1B_2B_3$ and $w = B_{\beta\beta}B_{\gamma\gamma} - B_{\beta\gamma}^2$, determine the volume element in the collective space. 

 The eigenvalue problem of the Hamiltonian~(\ref{eq1}) is solved using an expansion of eigenfunctions  
in terms of a complete set of basis functions that depend on five collective coordinates  $\beta, \gamma$ and $\Omega~(\phi,\theta,\psi)$ \cite{NLVPMR.09}. The eigenfunctions of the collective Hamiltonian read as
%
\begin{equation}
   \Psi^{IM}_\alpha(\beta, \gamma, \Omega)=\sum_{K\in\Delta I}\psi^I_{\alpha K}(\beta, \gamma)\Phi^I_{MK}(\Omega),
\end{equation}
For a given collective state, the probability density distribution in the $(\beta,\gamma)$ plane is defined as
\begin{equation}
\rho_{I\alpha}(\beta,\gamma) = \sum_{K \in \Delta I}{ \left| \psi_{\alpha K}^I(\beta,\gamma)\right|^2\beta^3 },
\label{eq:probability}
\end{equation}
with the summation over the allowed set of values of the projection $K$ of the angular  momentum $I$ on the body-fixed symmetry axis, and  with the normalization
\begin{equation}
\int_0^\infty{\beta d\beta \int_0^{2\pi}{ \rho_{I\alpha}(\beta,\gamma)|\sin{3\gamma}|d\gamma }}=1.
\end{equation}

 The reduced $E2$ transition probabilities are calculated by
\begin{eqnarray}
    B(E2;\alpha I\rightarrow\alpha'I') & = & 
    \sum_{\mu, M', M}  |\langle\alpha' I' M' |  \hat M(E2,\mu)| \alpha I M \rangle|^2  \nonumber \\
 &=&    \frac{1}{2I+1} | \langle\alpha'I'||\hat M(E2)||\alpha I\rangle|^2,
\end{eqnarray}
where $\hat M (E2, \mu)$ is the electric quadrupole operator which can be expressed 
in the following form \cite{KB.67}
\begin{equation}
\hat M (E2, \mu) = D_{\mu 0}^2 q_{20}^p(\beta, \gamma) +\frac{1}{\sqrt{2}}
(D_{\mu 2}^2 + D_{\mu -2}^2) q_{22}^p(\beta, \gamma)
\end{equation}
where
\begin{equation}
q_{2\kappa}^p = \langle \sum_{p}  e_p r_p^2 Y_{2\kappa}  \rangle
\end{equation}
are the quadrupole moments for protons at the deformation point $(\beta, \gamma)$ 
calculated in a fully self-consistent manner, the  indices $k$ equal to 0 and 2 and 
$e_p$ are the bare charges.

 In the framework of 5DCH-CDFT, the collective parameters of 5DCH, including the mass parameters $B_{\beta\beta}$, $B_{\beta\gamma}$, $B_{\gamma\gamma}$, the moments of inertia $\mathcal I_k$, and the collective potential $V_{\rm coll}$, are all determined microscopically from constrained triaxial CDFT calculations. The moments of inertia are calculated with Inglis-Belyaev formula \cite{inglis1956nuclear,beliaev1961concerning} and the mass parameters with the cranking approximation \cite{NLVPMR.09,GG.79}. The collective potential $V_{\rm coll}$ is calculated by
\begin{eqnarray}
V_{\rm coll}(\beta, \gamma) = E_{\rm tot}(\beta, \gamma) - \Delta 
V_{\rm vib}(\beta, \gamma) - \Delta V_{\rm rot} (\beta, \gamma), \nonumber \\
\label{V-coll}
\end{eqnarray}
where $E_{\rm tot}(\beta, \gamma)$ is the mean field total energy.
$\Delta V_{\rm vib}(\beta, \gamma)$ and $\Delta V_{\rm rot} (\beta, \gamma)$
are zero-point-energy (ZPE) values of vibrational and rotational motions. The collective 
ZPE corresponds to a superposition of zero-point  motion of individual nucleons in the 
single-nucleon potential. Here, the ZPE corrections are calculated in the cranking 
approximation \cite{NLVPMR.09,GG.79}.
     
   The energy surfaces  $E_{\rm tot}(\beta, \gamma)$ defined as a function of deformation 
parameters  $\beta$ and $\gamma$ are described as  potential energy surfaces (PES).
In the present manuscript, they are extracted from the triaxial relativistic mean field +BCS
(RMF+BCS) calculations.  The energy surfaces $V_{\rm coll}(\beta, \gamma)$ are 
labelled here as collective energy surfaces (CES); in addition to $E_{\rm tot}(\beta, \gamma)$
they contain  zero-point-energies of vibrational and rotational motion. The CES enter into
the action integral describing the fission dynamics (see Refs.\ \cite{KP.12,SR.16} and Eq.\ 
(\ref{action}) and its discussion below). Thus, the calculations of fission  fragment distributions, 
spontaneous fission half-lives etc depend sensitively on CES (see, for example, Refs.\ 
\cite{SNN.16,TZLNV.17} and references quoted therein).  In addition, the height of the fission barrier is defined as an 
energy difference between the saddle point and minimum of CES, namely, as 
$V_{coll}(saddle) - V_{coll}(min)$  (see, for example, Refs.\ \cite{ELLMR.12,RR.14,GPR.18}).
Note that different approaches exist for the calculations of ZPE contributions to CES  and in a 
number of publications  zero-point-energies of vibrational motion are neglected since their 
variation with deformation is rather modest (see, for example, Ref.\ \cite{RR.14}).

\begin{figure}[h!]
\centering
\includegraphics[scale=0.6,angle=0]{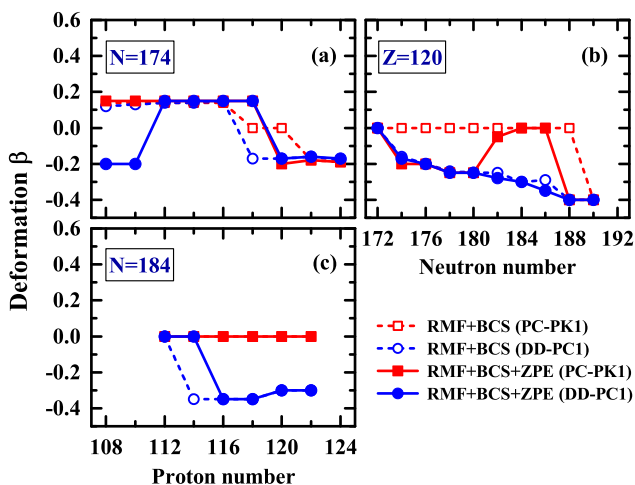}
\caption{(Color online)  The quadrupole deformations of the
minima in potential  and collective energy surfaces obtained in
the RMF+BCS and  RMF+BCS+ZPE calculations, respectively. The results
obtained with DD-PC1 and PC-PK1 CEDFs are presented for the $N=174$
[panel(a)] and $N=184$ [panel(c)] isotonic chains as well as for
the $Z=120$ isotopic chain [panel(b)].}
\label{deformation_GS_-ddpc1-pcpk1}
\end{figure}

\section{Results and Discussion}\label{sec2}

   The starting point is triaxial RMF+BCS calculations. These 
calculations are performed imposing constraints on the axial $Q_{20}$ and triaxial $Q_{22}$ 
mass quadrupole moments.  Note, that full-scale calculations are performed on the grid 
which covers the quadrupole deformation range $\beta_2=0 - 0.6$ in steps of $\Delta \beta_2 = 0.05$ 
and gamma deformation range $\gamma = 0^{\circ} - 60^{\circ}$ in steps of
$\Delta \gamma = 6 ^{\circ}$.

  In order to avoid the uncertainties connected with the definition of
the size of the pairing window \cite{KALR.10}, we use the separable form
of the finite range Gogny pairing interaction introduced by Tian et al
in Ref.\ \cite{TMR.09} which, in addition, is multiplied by scaling factor
$f$ (see Eq.\ (25) in Ref.\ \cite{AARR.14}). The systematic investigation 
of pairing properties in the actinides
\cite{AARR.14,AO.13} indicates that scaling factor $f=1.0$ is appropriate
for the relativistic Hartree-Bogoliubov (RHB) description of actinides and
superheavy nuclei and different physical observables in these mass
regions are well reproduced with such a factor
\cite{AANR.15,A250,AARR.14,AAR.16,AO.13}.
However, the experience shows that this factor has to be larger in the RMF+BCS
framework as compared with the RHB one \cite{XLYLRM.13}. Thus, for the RMF+BCS framework this
factor has been defined by matching the gain of binding due to proton and neutron
pairing (as compared with unpaired solution) obtained in the axial RHB
calculations with $f_\nu=f_\pi=1.0$ for the $-0.6 < \beta_2 < 0.6$ deformation
range of the $^{308}120$ nucleus. This led  to the following neutron and proton
scaling factors for the RMF+BCS calculations: $f_\nu=1.066$ and $f_\pi=1.052$ for
the DD-PC1 functional and  $f_\nu=1.073$ and $f_\pi=1.058$ for the PC-PK1 CEDF.

   The truncation of the basis is performed in such a way that all states belonging to 
the major shells up to $N_F=18$ fermionic shells are taken into account for the Dirac 
spinors. This truncation of the basis provides excellent numerical accuracy for the 
ground state properties and sufficient numerical accuracy for the changes of fission 
barrier heights due to the correlations beyond mean field. This basis is also sufficient 
for the calculation of the $E(2_1^+)$ energies and $B(E2; 2_1^+ \rightarrow  0_1^+)$  
transition probabilities. This was verified by comparing the results of the 5DCH 
calculations with $N_F=18$ and $N_F=20$ for a few nuclei; the $N_F=18$ and $N_F=20$ 
results for these observables differ marginally and cannot be discriminated on the plots 
presented in Figs.\ \ref{E21+energies} and \ref{BE2-values}  below.

  On the other hand, numerically accurate calculations of absolute values of fission 
 barrier heights  in this mass region require the fermionic basis with $N_F=20$ (see 
 Refs.\ \cite{AARR.17,AAR.12}). However, the 5DCH calculations in such a basis 
 are prohibitively expensive and they have to be performed at all grid points.  To 
 overcome this problem we use the fact that the dynamical contributions to fission 
 barrier height, defined as
 \begin{eqnarray}
 E^{FB}_{dyn} & = & [\Delta V_{\rm vib}(\beta, \gamma) + \Delta V_{\rm rot} (\beta, \gamma)]_{saddle} 
 \nonumber \\ 
 && - [\Delta V_{\rm vib}(\beta, \gamma) + \Delta V_{\rm rot} (\beta, \gamma)]_{ground\,\, state}, 
 \label{dyn_FB}
\end{eqnarray}
calculated with $N_F=18$ and $N_F=20$ differ by less than 20 keV.
This clearly indicates that numerical errors in the fission barriers are dominated by the 
truncation errors in the mean field part. This result is born in few detailed full-scale calculations.

  Note that the topologies\footnote{The topology of potential energy surface means the general shape 
of multidimensional potential energy surface in terms of minima and saddles and general 
connectivity that characterize such a surface \cite{BK.97}.} of potential energy surfaces (PES) 
[collective energy surfaces (CES)]  are very similar in the calculations with $N_F=18$ and $N_F=20$
with the deformations of the ground states and saddles being almost independent of $N_F$.
This allows simplified approach discussed below to the calculation of the fission barriers in the RMF+BCS and 
 RMF+BCS+ZPE calculations. First, based on PES and CES obtained in the calculations with $N_F=18$, we 
 define the regions  close to the saddle of fission barrier and ground state. Second, for these regions, 
 the RMF+BCS calculations are repeated with $N_F=20$ and RMF+BCS fission barrier $E^{FB}_{\rm RMF+BCS}
 (N_F=20)$ is defined for $N_F=20$. Such procedure has been used earlier in Ref.\ \cite{AARR.17}. 
 Third, the fission barrier in CES is defined as 
 \begin{eqnarray}
 E^{FB}_{\rm RMF+BCS+ZPE} (N_F=20) & = & E^{FB}_{\rm RMF+BCS+ZPE} (N_F=18) \nonumber \\
+ [E^{FB}_{\rm RMF+BCS}
 (N_F=20) & - & E^{FB}_{\rm RMF+BCS}
 (N_F=18)]
 \end{eqnarray}
This procedure saves a lot of computational time since
the $N_F=20$ calculations are performed only on a limited part of the grid space
and they are performed only at the mean field level.

\begin{figure}[htb]
\centering
\includegraphics[scale=0.6,angle=0]{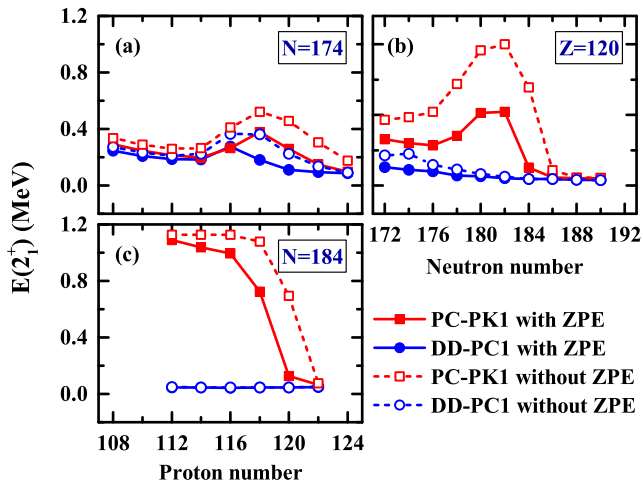}
 \caption{(Color online) Excitation energies of the $2_1^+$ states
a function of proton number in the $N=174$ and 184 isotonic
chains and as a function of neutron number in the $Z=120$ isotopic
chain.}
\label{E21+energies}
\end{figure}

\begin{figure}[htb]
\centering
\includegraphics[scale=0.6,angle=0]{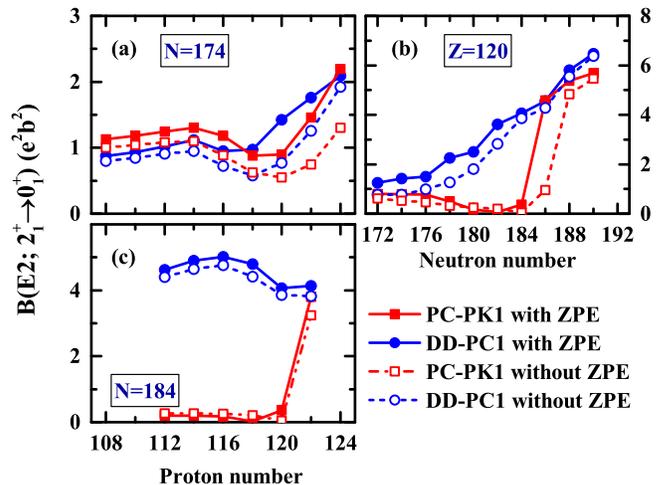}
\caption{(Color online) The $B(E2; 2_1^+ \rightarrow 0_1^+)$
values as a function of proton number in the $N=174$ and 184 isotonic
chains and as a function of neutron number in the $Z=120$ isotopic
chain.}
\label{BE2-values}
\end{figure}

\begin{figure*}[htb]
\centering
\includegraphics[scale=0.8,angle=0]{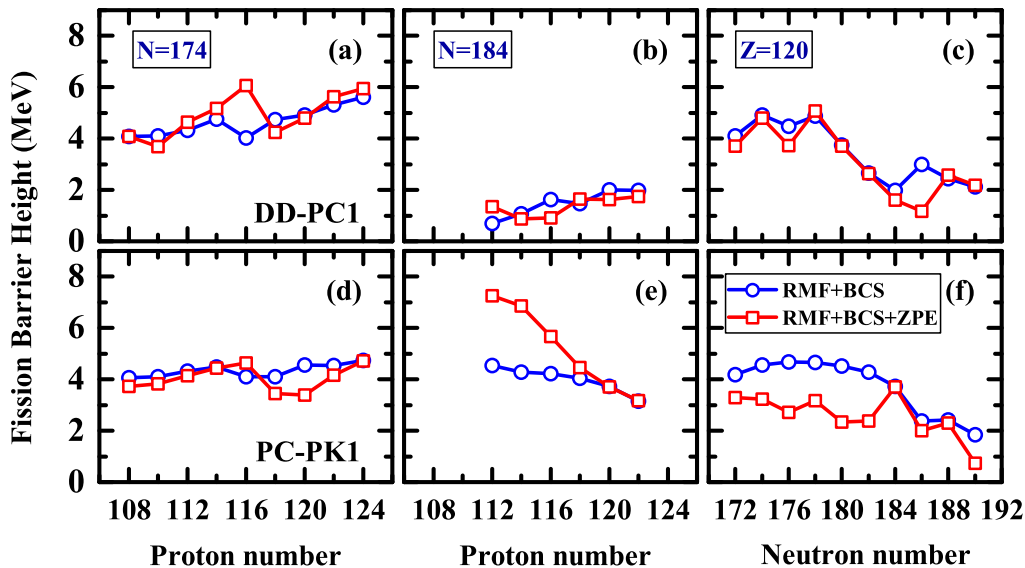}
\caption{(Color online) The heights of inner fission barriers in the mean
field calculations (labeled as 'RMF+BCS') and in the calculations with
dynamical correlations included (labeled as 'RMF+BCS+ZPE').}
\label{full_barrier}
\end{figure*}

   The potential energy surfaces  for the $Z=120$ isotopes with
$N=172$, 178, 184 and 190 obtained in the RMF+BCS calculations
with the PC-PK1 functional are shown in Fig.\ \ref{pes-z=120-pcpk1}.
The minima are located at spherical shape for $N=172$, 178, and 184
and only the $^{310}$120 nucleus has an oblate ground state with
$\beta_2 \sim -0.4$. Note that PES are soft in quadrupole deformation 
in the vicinity of the minima. As a result, the inclusion of ZPE leads 
to substantial modifications in a number of nuclei. For example, the 
$N=178$ nucleus is no longer spherical in its ground state since the 
minimum in collective energy surface is located at $\beta_2 \sim -0.25$ 
(see Fig.\ \ref{pes-z=120-pcpk1}). In addition, the collective energy 
surfaces are very soft in quadrupole deformation.  As a consequence, 
the wavefunction of the $^{292}$120 nucleus is localized at 
$\beta_2 \sim -0.15$ despite the fact the minimum in collective energy 
is located at spherical shape. Thus, contrary to previous mean field 
studies this nucleus cannot be considered in the 5DCH calculations as 
``doubly magic'' spherical nucleus.

 Fig.\ \ref{deformation_GS_-ddpc1-pcpk1} summarizes the results
for the deformations of the minima in potential and collective energy
surfaces for the $Z=120$ isotopic and $N=174, 184$ isotonic chains
obtained in the RMF+BCS and RMF+BCS+ZPE calculations with the PC-PK1 
and DD-PC1 CEDFs. These surfaces as well as probability densities
distributions for the $0^+_1$ collective wavefunctions are
presented in Figs. 1-18 of supplemental material. One can see
substantial changes in equilibrium deformation of the nuclei
located in transitional regions when the correlations beyond
mean field are included. For example, the transition from
prolate to oblate shape is triggered in the $N=174$ nuclei
with $Z=108$ and 110 when ZPE are included in the calculations
with DD-PC1 (Fig.\ \ref{deformation_GS_-ddpc1-pcpk1}a). ZPE
also triggers the transition from spherical shape to deformed
one in the $N=174$ nuclei with $Z=118$ and 120 in the
calculations with PC-PK1. The modifications are smaller in the
$N=184$ isotonic chain (Fig.\ \ref{deformation_GS_-ddpc1-pcpk1}c);
the deformation of the energy minimum is switched from highly
deformed oblate to spherical one only in the $Z=114$ nucleus
in the calculations with DD-PC1 when ZPE are added. Otherwise, the
DD-PC1 and PC-PK1 CEDF give distinctly different predictions
for the deformations of the CES minima in the $N=184$ nuclei.
The  former functional predicts mostly oblate shapes in the
ground state, while the latter one only spherical shapes. The
deformations of the $Z=120$ nuclei are very weakly affected by
the ZPE's in the calculations with DD-PC1 (Fig.\
\ref{deformation_GS_-ddpc1-pcpk1}). On the contrary, they are
drastically affected by ZPE in the case of PC-PK1 CEDF; the
deformations of the minima of the $N=174-180$ and $N=188$
isotopes change from spherical to oblate ones when ZPE is
added.  While the results of the RMF+BCS calculations for $Z=120$
nuclei are drastically different for the PC-PK1 and DD-PC1,
they mostly
converge
to the same deformed oblate solution when ZPE are added.
This points to reduced role of the $Z=120$ proton shell gap
which in many earlier RMF studies was interpreted as ``magic''
one.

\begin{figure}[h!]
\centering
\includegraphics[scale=0.6,angle=0]{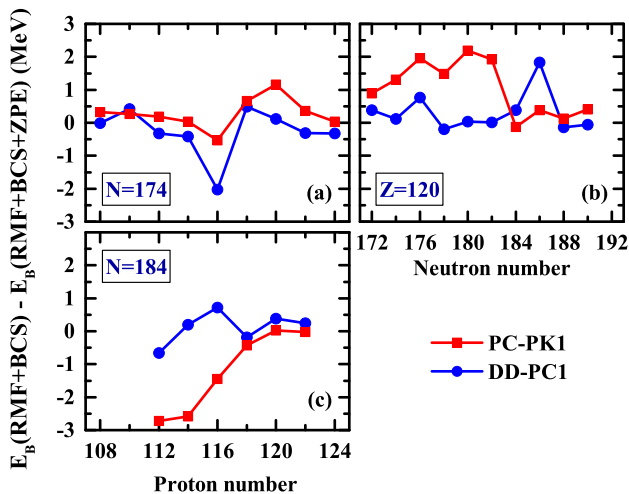}
\caption{(Color online) The same as Fig.\ \ref{deformation_GS_-ddpc1-pcpk1} 
but for the impact of dynamical correlations on the height of inner fission
barrier. Negative (positive) value of E$_{\rm B}$(RMF+BCS)-E$_{\rm B}$(RMF+BCS+ZPE)
means higher (lower) fission barrier in the calculations with dynamical correlations
included.}
\label{barrier-ddpc1-pcpk1}
\end{figure}

\begin{figure*}[htb]
\centering
\includegraphics[scale=0.8,angle=0]{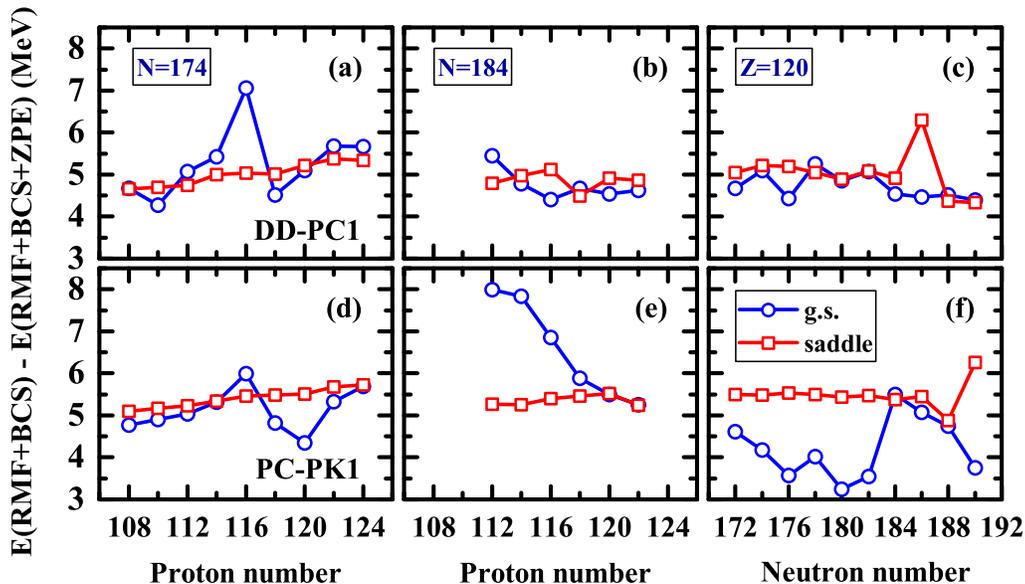}
\caption{(Color online) Dynamical correlations energies at the
ground states and the saddles of inner fission barriers of the 
nuclei under study.}
\label{dyn_energies}
\end{figure*}

  Further information on the collectivity of the states of interest
can be obtained by analysing $E(2^+_1)$ energies (Fig.\ \ref{E21+energies})
and the $B(E2; 2_1^+  \rightarrow 0_1^+)$ transition rates (Fig.\
\ref{BE2-values}). The excitation energies and transition rates are 
strongly affected  by the quadrupole deformations of the respective minima and also by the 
dynamics of large shape fluctuations around equilibrium shape which 
strongly depends on the topology of PES. These properties can be 
reasonably well described in the 5DCH calculations 
as illustrated by the studies of the Sn isotopes in Ref.\ \cite{LLXYM.12}.

  With the exception of the $Z=118$ and 120 nuclei, the results for
the $E(2^+_1)$ values are very similar for the $N=174$ isotones in the
calculations with PC-PK1 and DD-PC1 CEDFs (Fig.\ \ref{E21+energies}a).
Substantial difference between the $B(E2; 2_1^+  \rightarrow 0_1^+)$
values obtained with these two functionals is observed only at $Z=120$
(Fig.\ \ref{BE2-values}a). Note that in this isotonic chain spherical
shapes appear on the mean field level only for the $Z=118$ and 120
nuclei in PC-PK1 CEDF (Fig.\ \ref{deformation_GS_-ddpc1-pcpk1}); this
is a reason for some weakening of the collectivity in these nuclei in
the 5DCH calculations with PC-PK1 as compared with the ones based on
DD-PC1 CEDF.

  In the $Z=120$ isotopic chain, the $N=180-184$ nuclei are
significantly less collective in the calculations with CEDF PC-PK1
as compared with DD-PC1 (Figs.\ \ref{E21+energies}b and
\ref{BE2-values}b). Above $N=186$, there is no difference
between the DD-PC1 and PC-PK1 results. Below $N=178$, the nuclei
are less collective in the calculations with PC-PK1 but the difference
is not that significant as in the $N=180-184$ nuclei. All these
features closely correlate with the
presence/absence of spherical nuclei along
the $Z=120$ chain in the RMF+BCS calculations with PC-PK1/DD-PC1
functionals and with the modifications of PES induced by dynamical correlations
(see Fig.\ \ref{deformation_GS_-ddpc1-pcpk1}b and Fig.\ \ref{pes-z=120-pcpk1}).

  The results for the $N=184$ nuclei with $Z=112-118$ obtained with
the DD-PC1 and PC-PK1 functionals are distinctly different (Figs.\
\ref{E21+energies}c and \ref{BE2-values}c). Indeed, the combination
of $E(2^+_1) \sim 1.0$ MeV (which is substantially higher than the
$E(2^+_1)$ values obtained for the $N=174$ and $Z=120$ chains) and low
$B(E2; 2_1^+  \rightarrow 0_1^+)$ values obtained in the 5DCH calculations
with PC-PK1 strongly suggests that the $Z=112-118$, $N=184$ nuclei may be
considered as truly spherical. The $Z=120$, $N=184$ nucleus is transitional 
in nature with small $E(2^+_1)$ and $B(E2; 2_1^+  \rightarrow 0_1^+)$ values
and the $Z=122$, $N=184$ nucleus is collective in ground state in the 
calculations with PC-PK1. On the other hand, all $N=184$ isotones are 
collective in their ground states in the calculations with DD-PC1.

  It is interesting to investigate the impact of ZPE on the  $E(2^+_1)$ energies      
and $B(E2; 2_1^+  \rightarrow 0_1^+)$ transition rates. This is done by neglecting
ZPE in the 5DCH calculations; such results are shown by dashed lines with open
symbols in Figs.\ \ref{E21+energies} and \ref{BE2-values}.

 The neglect of ZPE typically leads to the increase of the $E(2^+_1)$ energies 
and there is a correlation between the magnitude of this increase and the 
deformation of the system.  This 
increase is either small or even non-existent in the results obtained with the DD-PC1 
functional (see Fig.\ \ref{E21+energies}); note that the absolute majority of the 
calculated nuclei have non-zero  $\beta$ deformations for the ground states in the 
RMF+BCS and 5DCH calculations with this functional (see Fig.\ 
\ref{deformation_GS_-ddpc1-pcpk1}). These deformations are similar or have similar 
magnitude for the $N=174$ isotopic chain in the calculations with DD-PC1 and 
PC-PK1 (see Fig.\  \ref{deformation_GS_-ddpc1-pcpk1}a).  As a result, the increase 
in the $E(2^+_1)$ energies due to neglect of ZPE is comparable in both functionals.
On the contrary, in the calculations with PC-PK1 the increases in the $E(2^+_1)$ energies due to neglect of ZPE are
larger in the $N=184$ isotonic chain (see Fig.\ \ref{E21+energies}c) and they are 
especially large in the $Z=120$ isotopic chain (see Fig.\ \ref{E21+energies}b). In
the former chain, the ground states of the nuclei have $\beta=0$ both in the RMF+BCS and RMF+BCS+ZPE
calculations (see Fig.\ \ref{deformation_GS_-ddpc1-pcpk1}c). In the latter chain, 
with exception of the $N=190$ nucleus, in the ground states the $\beta$ deformation  
is zero in the RMF+BCS calculations and the inclusion of ZPE triggers the transition 
to oblate deformation in the nuclei with $N=174-180$ and $N=188$ (see 
Fig.\ \ref{deformation_GS_-ddpc1-pcpk1}b).

  In general, the neglect of ZPE leads to the decrease of the 
$B(E2; 2_1^+  \rightarrow 0_1^+)$ transition rates (see Fig.\ \ref{BE2-values}).
The only exceptions are the $N=184$ nuclei with $Z=112-118$ (see Fig.\ 
\ref{BE2-values}c) and $Z=120$ nuclei with $N=180,182$ (see Fig.\ 
\ref{BE2-values}b) in the calculations with PC-PK1. However, these nuclei 
are characterized by very low values of  $B(E2; 2_1^+  \rightarrow 0_1^+)$. 
Note also that the decrease of the  $B(E2; 2_1^+  \rightarrow 0_1^+)$ transition 
rates due to the neglect of ZPE depends on the nucleus and on the functional. 
Note that no direct correlations between these decreases in the
$B(E2; 2_1^+  \rightarrow 0_1^+)$ values and the topologies of PES and/or 
CES of the nuclei under consideration have been found.

  Fig.\ \ref{full_barrier} shows the impact of dynamical correlations on 
the heights of inner fission barriers of the nuclei under consideration.
 In the mean field calculations, the height of fission barrier is defined
as the energy difference between the saddle point and minimum of PES.
In the beyond mean field calculations, this energy difference is extracted 
from the energies of saddle and minimum of collective energy surface: 
this is a consistent with the definition of fission barrier height used 
before  in beyond mean field approaches based on Gogny and Skyrme 
energy density functionals \cite{ELLMR.12,RR.14,GPR.18}.

 The changes introduced in the fission 
barrier heights due to dynamical correlations are summarized in Fig.\ \ref{barrier-ddpc1-pcpk1}.
The calculated heights obtained in the RMF+BCS calculations
are in general close  to the ones obtained in the RHB calculations of Ref.\ 
\cite{AARR.17}; some differences are due to the use of different frameworks (RMF+BCS in the 
present manuscript and RHB in Ref.\  \cite{AARR.17}) and the differences in the way the pairing 
interaction has been defined in both manuscripts. Note that the height of fission barrier extremely 
sensitively depends on the  strengths of pairing interaction (see Ref.\ \cite{KALR.10} and references 
quoted therein).

 One can see that in the calculations with the DD-PC1 functional, the fission barriers obtained 
 in the calculations with and without dynamical correlations are close to each other;  
the modifications of the fission barrier height by the dynamical correlations are typically in the 
range of $\pm 0.5$ MeV (see Fig.\ \ref{barrier-ddpc1-pcpk1}). The only exceptions are the 
$(Z=116, N=174)$ (Fig.\ \ref{barrier-ddpc1-pcpk1}a), $(Z=120,N=186)$  
(Fig.\ \ref{barrier-ddpc1-pcpk1}b) and $(Z=116, N=184)$ (Fig.\
\ref{barrier-ddpc1-pcpk1}c) nuclei for which the modifications of inner fission barrier due
to dynamical correlations are close to or exceed 1 MeV. Note that the absolute majority of  the nuclei under 
consideration are deformed in the ground states in the calculations at and beyond mean field 
levels with DD-PC1 functional (see Fig.\ \ref{deformation_GS_-ddpc1-pcpk1} and Figs. 4, 5, 10, 
11, 16 and 17 in supplemental material). 

  Similar features are also seen for the $N=174$ isotones in the calculations with the PC-PK1 
functional (see Fig.\ \ref{full_barrier}d). The majority of the nuclei in the $N=174$ chain are 
deformed both at and beyond mean field levels (see Fig.\ \ref{barrier-ddpc1-pcpk1}a and Figs. 7 
and 8 in supplemental material) and only $Z=118$ and 120 nuclei are spherical in the mean field 
calculations. Only for the latter two nuclei the modifications of the fission barrier height by 
dynamical correlations are close to or exceed 1 MeV (see Fig.\ \ref{barrier-ddpc1-pcpk1}a).

 On the contrary, substantial changes in fission barrier heights induced by dynamical correlations 
are  seen in the nuclei which are spherical in the ground states in the RMF+BCS calculations with
PC-PK1.  These are the $Z=118, 120$ nuclei in the $N=174$ isotopic chain, the $N=172-182$ nuclei 
in the $Z=120$ chain and the $Z=112-116$ nuclei in the $N=184$ chain. Dynamical correlations lead 
to a substantial increase (decrease) of fission barriers in the $N=184$ isotones with $Z=112-116$ 
(in the $Z=120$ isotopes with $N=172-182$). However, they have very limited impact of the fission 
barriers of spherical nuclei located in close vicinity of the $Z=120, N=184$ nucleus; these are nuclei
which have the features of spherical nucleus both at and beyond  mean field levels.
 
  To better understand the origin of these changes in the fission barrier heights we plot 
dynamical correlation energies for the ground states and the saddles of inner fission barriers 
in Fig.\ \ref{dyn_energies}. Several interesting features emerge from the analysis of this
figure. First, the variation of dynamical correlation energies with neutron number is rather smooth
at the saddles of inner fission barriers. Moreover, these energies are around of 5 MeV
in all nuclei under study. On the contrary, dynamical correlation energies for the ground states typically 
show  much larger fluctuations as a function of neutron number; these fluctuations are 
especially pronounced for the chains of the nuclei which are calculated to be spherical at the
mean field level. Second, these dynamical correlation energies are very similar at the 
ground state and the saddle of inner fission barrier in deformed nuclei (see  Fig.\ 
\ref{dyn_energies}a, b,  c and d). As a consequence, the impact of dynamical correlations 
on the fission barriers of deformed nuclei is limited.  On the other hand, they are quite
different in the nuclei which have spherical ground states in the mean field 
calculations. This feature explains observed increase of the importance of dynamical 
correlations for the calculation of inner fission barrier of SHE with soft PES the minimum
of which is located at spherical shape.

\begin{figure*}[htb]
\centering
\includegraphics[scale=0.8,angle=0]{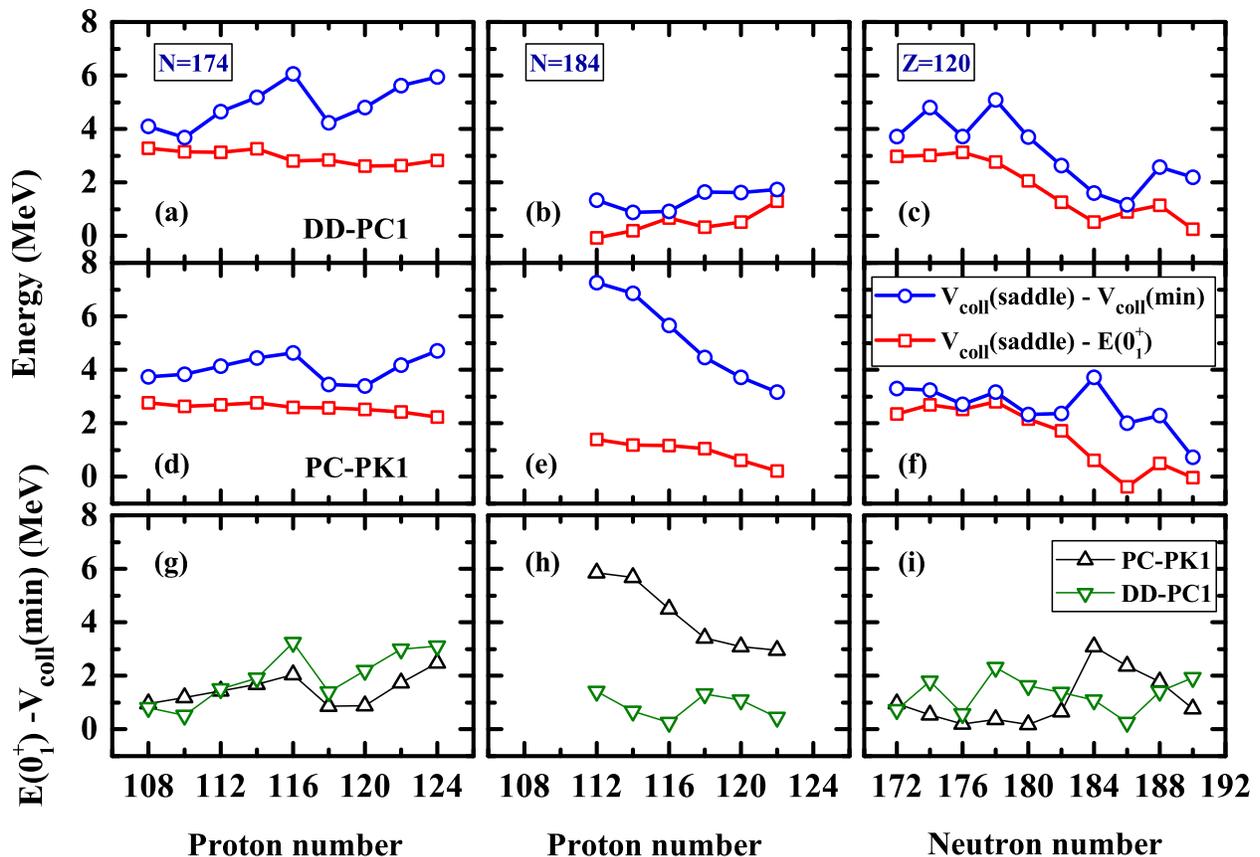}
\caption{The  energy difference $V_{coll}(saddle) - V_{coll}(min)$ between the saddle 
point and the minimum of collective energy surface compared with the energy difference 
$V_{coll} - E(0^+_1)$. The energy $E(0^+_1) -V_{coll}(min)$ (shown in bottom panels) is 
the ground state energy (relatively to the energy of the  minimum of CES) in the 5DCH 
calculations.}
\label{Fig-extra}
\end{figure*}

  It is also interesting to look on potential impact of the ground state energy on the description of some fission 
processes. For 
example, the calculation of spontaneous fission half lives $\tau_{SF}$ depends on the energy $E$ of collective 
ground state (see Ref.\ \cite{BKRRSW.15}) since it enters into the action integral $S$, corresponding to trajectory 
between two points {\bf a} and {\bf b} in {\bf q}-space  (collective coordinate space),
\begin{equation}
S({\bf a}, {\bf b}, E) = \int_{0}^{s} \sqrt{2B_s ({\bf q}(s')) [E-V({\bf q}(s'))]} ds'  
\label{action}
\end{equation}
where the trajectory length counts from zero at {\bf a} to $s$ at {\bf b} (see Sect. 5.1.3. in Ref.\ \cite{KP.12}). 
In many applications, the tunneling energy $E$ (which is also the ground state energy of the nucleus before 
fission) is either approximated by $E_0=0.5$ MeV (see Refs.\ \cite{NTSSWGLMN.69,WERP.02,BLSK.05,WE.12,RR.14}) 
or defined from WKB quantization rules (see Ref.\  \cite{BKRRSW.15}). In the latter case, this  energy is extracted 
from  the condition that $V(q)=E_0$ at classical turning points.

  On the contrary, one can take a more microscopic approach and associate tunneling energy $E$ 
with the energy of collective ground state defined either in Generator Coordinate Method (GCM) or in 
5DCH.  To our knowledge, this has been done so far only in Ref.\ \cite{SEKRH.09} in which the collective 
ground state energy is defined from GCM calculations; these calculations are based on Skyrme energy density 
functional but are restricted to axial shape. As discussed in Ref.\ \cite{BKRRSW.15}, the microscopic values of 
tunneling energies differ from approximate ones. In a similar fashion, one can associate the tunneling energy 
$E$  with the energy $E(0^+_1)$ of the ground state obtained in 5DCH.  The  $E(0^+_1)$ energies, shown
in bottom panels of Fig.\ \ref{Fig-extra},
deviate substantially in many cases both from $E_0=0.5$ MeV and 
from the ground state energies defined by means of the WKB quantization rules (which are displayed in Fig. 4 
of Ref.\ \cite{BKRRSW.15}). The differences between these values also substantially depend on proton and 
neutron numbers.

   These differences in the values of tunneling energy $E$ are expected to have a profound effect on 
spontaneous fission half lives $\tau_{SF}$. Although the calculation of  $\tau_{SF}$ is beyond the scope 
of the present manuscript, the comparison of $V_{coll}(saddle) - E(0_1^+)$  and  $V_{coll}(saddle) - V_{coll}(min)$
allows to estimate the major trends. The difference $V_{coll}(saddle) - E(0_1^+)$ defines the 
maximum variation of the $(E-V({\bf q}))$ difference in the action integral of Eq.\ \ref{action}. It is lower than the fission 
barrier height $V_{coll}(saddle) - V_{coll}(min)$ typically by more than 0.5 MeV. In many cases this
difference reaches few MeV. All this suggest that the approximation of tunneling energy by 
$E_0=0.5$ MeV  (as done in many applications) is highly unreliable. The present results 
suggest that the use of  $E(0_1^+)$  (as defined by 5DCH) for the energy of collective
ground state will result in a  substantial reduction of spontaneous fission half-lives as
compared with estimates based on $E=E_0$.
  
  The magnitude of the $E(0^+_1)$ with respect of the minimum of the collective energy surface
(the  $E(0^+_1) - V_{coll}(min)$ quantity in Fig.\ \ref{Fig-extra}) depends on the softness of
collective energy surface in the vicinity of spherical/normal-deformed minimum. The soft (stiff)
CES leads to low (high) values of the $E(0^+_1) - V_{coll}(min)$ quantity.  This dependence is
especially pronounced in the $N=184$ isotonic chain (see Fig.\ \ref{Fig-extra}h). The CES's of these
isotopes are soft in the vicinity of spherical minimum at $Z=112$ and oblate minimum at $Z=114-122$
in the DD-PC1 functional (see Fig.\ 17 in supplemental  material) and this leads to  $E(0^+_1) - V_{coll}(min) 
\approx 1.0$ MeV. On the contrary, the CES's are stiffer in the vicinity of spherical minimum in the PC-PK1 
functional with their stiffness decreasing with increasing $Z$ (see Fig.\ 14 in supplemental material) and this 
leads  to substantially higher $E(0^+_1) - V_{coll}(min)$ values which decrease with increasing $Z$
(see Fig.\ \ref{Fig-extra}h). Similar correlations between the softness of CES in the vicinity of the minimum 
under consideration  and the $E(0^+_1) - V_{coll}(min)$ values can be found in the $Z=120$ isotopic chain 
(compare Figs.  2 and 5 in supplemental material with Fig.\ \ref{Fig-extra}i in the manuscript) and $N=174$ 
isotonic chain (compare Figs.  8 and 11 in supplemental material with Fig.\ \ref{Fig-extra}g).

\section{Summary}\label{sec3}

  In conclusion, the impact of beyond mean field effects on the
ground state and fission properties of superheavy nuclei has
been investigated in five-dimensional collective Hamiltonian. We focus
here on two functionals (DD-PC1 and PC-PK1) which give distinctly
different predictions along the $Z=120$ and $N=184$ lines at the mean
field level. For the first time it is shown that the inclusion of
dynamical correlations brings the predictions of these two functionals
closer for nuclei along the $Z=120$ line. Only few nuclei around $N=184$
remain spherical in the calculations with PC-PK1; the rest of nuclei
possess significant collectivity. This stresses again that the impact of
spherical shell closure at $Z=120$ is quite limited. On the contrary,
the predictions of these two functionals remain distinctly
different for the $N=184$ nuclei even when dynamical correlations
are included. These nuclei are mostly spherical (oblate) in the
calculations with PC-PK1 (DD-PC1).  The impact of dynamical correlations
on the height of inner fission barrier has been investigated. It is
typically moderate (significant) when the ground state is deformed
(spherical) at the mean field level. This result for the first time
shows the importance of the inclusion of dynamical correlations for the
calculation of inner fission barriers of the superheavy nuclei with
soft potential energy surfaces the minimum of which at mean field level
is located at spherical shape.

  It is important to keep in mind that potential energy surfaces of many 
superheavy nuclei are soft also in non-relativistic theories (see, for example, Refs.\ 
\cite{CHN.05,JKS.11,WE.12}). It is reasonable to expect that similar 
to this study the correlations beyond mean field could have a substantial 
impact on their ground state and fission properties and potentially on 
the localization and the properties of predicted islands of stability 
of superheavy elements.

\begin{acknowledgments}
  This material is based upon work supported by the US Department of
Energy, Office of Science, Office of Nuclear Physics under Award No.
DE-SC0013037, by the CUSTIPEN (China-U.S. Theory Institute for Physics 
with Exotic Nuclei) funded by the U.S.  Department of Energy, Office of 
Science under grant number DE-SC0009971 and by the National Key R\&D
Program of China (Contract No. 2018YFA0404400) and the National Natural
Science Foundation of China (Grants No. 11335002, No. 11475140, No. 
11575148, No. 11621131001, and No. 11875225).
\end{acknowledgments}

\end{document}